\documentclass{PoS}

\title{Probing heavy ion collisions using quark and gluon jet substructure}

\ShortTitle{Quark and gluon jet substructure in heavy ion collisions}

\author{\speaker{Yang-Ting Chien}\thanks{The author would like to thank the Hard Probes 2018 organizers for the hospitality and support.}\\
        Center for Theoretical Physics, Massachusetts Institute of Technology, Cambridge, MA 02139, U.S.A.\\
        E-mail: \email{ytchien@mit.edu}}

\abstract{Understanding the inner working of the quark-gluon plasma requires complete and precise jet substructure studies in heavy ion collisions.
In this proceeding we discuss the use of quark and gluon jets as independent probes, and how their classification allows us to uncover regions of QCD phase space sensitive to medium dynamics. We introduce the telescoping deconstruction (TD) framework to capture complete jet information and show that TD observables reveal fundamental properties of quark and gluon jets and their modifications in the medium. We draw connections to soft-drop subjet distributions which help illuminate medium-induced jet modifications. The classification is also studied using a physics-motivated, multivariate analysis of jet substructure observables. Moreover, we apply image-recognition techniques by training a deep convolutional neural network on jet images to benchmark classification performances. We find that the quark-gluon discrimination performance worsens in \textsc{Jewel}-simulated heavy ion collisions due to significant soft radiation affecting soft jet substructures. This work suggests a systematic framework for jet studies and facilitates direct comparisons between theoretical calculations and measurements in heavy ion collisions.}

\FullConference{International Conference on Hard and Electromagnetic Probes of High-Energy Nuclear Collisions\\
		30 September - 5 October 2018\\
		Aix-Les-Bains, Savoie, France}

\begin{document}

\begin{figure}
    \includegraphics[height=2.4cm, trim = {0mm 0mm 0mm 0cm}, clip]{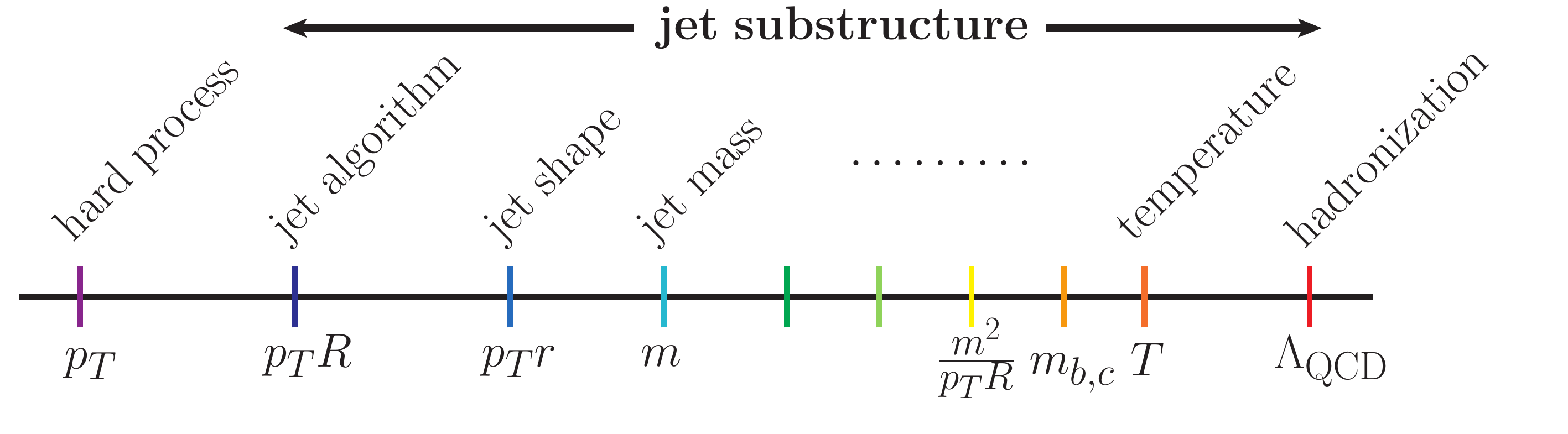}
    \includegraphics[height=2.6cm, trim = 0mm 0cm 0mm 0cm , clip=true]{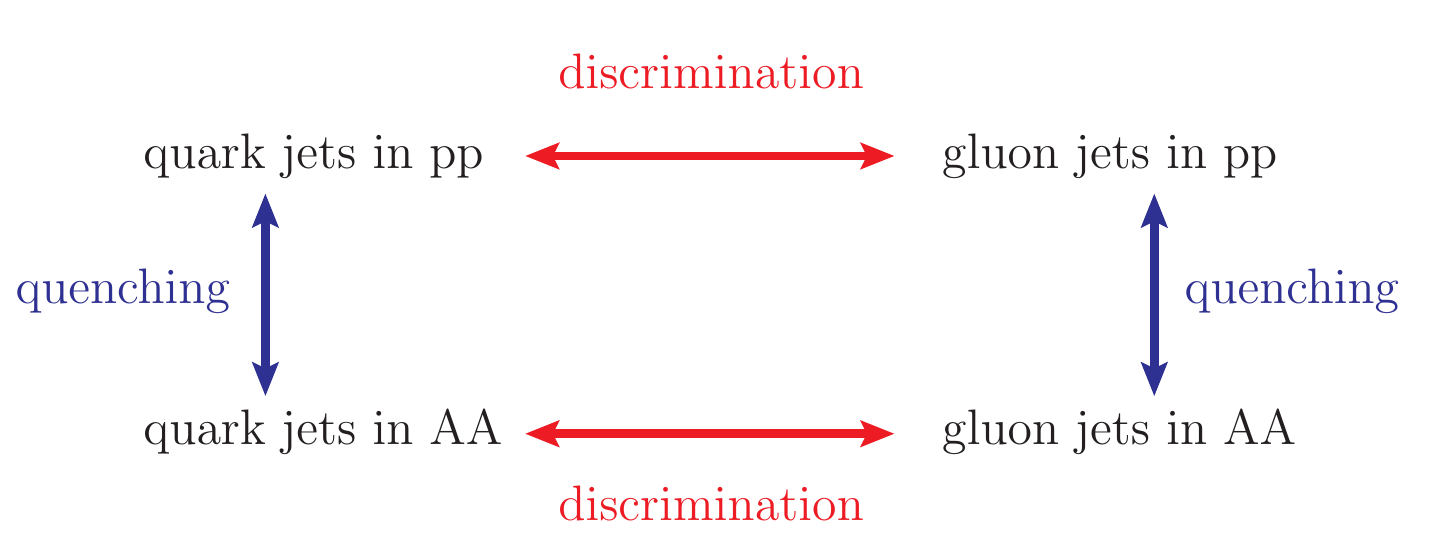}
\caption{Left panel: Jet substructure observables can probe QCD dynamics in all energy regimes from the highest scale down to $\Lambda_{\rm QCD}$ \cite{Chien:2015hda}. Right panel: Classification of quark and gluon jets in pp and AA collisions provide a new method for jet quenching studies \cite{Chien:2018dfn}.}
    \label{fig:sub_scale}
\end{figure}

\section{Introduction}
\label{intro}

The study of jet quenching has moved onto detailed analysis of the redistribution of jet energy quantified by jet substructure modifications. Different jet substructure observables are sensitive to different underlying QCD dynamics at characteristic energy scales (left panel of Fig. \ref{fig:sub_scale}). One can design jet substructure observables to probe specific regions of phase space where jet-medium interaction may have the dominant effect. A comprehensive examination of jet substructure modifications will then allow us to search for possible signatures which may reveal fundamental properties of the quark-gluon plasma (QGP). On the other hand, a change of quark and gluon jet fractions in heavy ion collision can contribute significantly to jet substructure modifications. An increase of the quark-jet fraction due to larger suppressions of gluon jets can make the jet energy profile more quark-jet like, an important effect in addition to the jet-by-jet modification to substructure \cite{Chien:2015hda,Chatrchyan:2013kwa}. This further motivates the studies of jet modifications with different quark and gluon jet fractions which enable the use of quark and gluon jets as independent probes.

\begin{figure}
    \includegraphics[height=4.23cm, trim = {0mm 0mm 0mm 0cm}, clip]{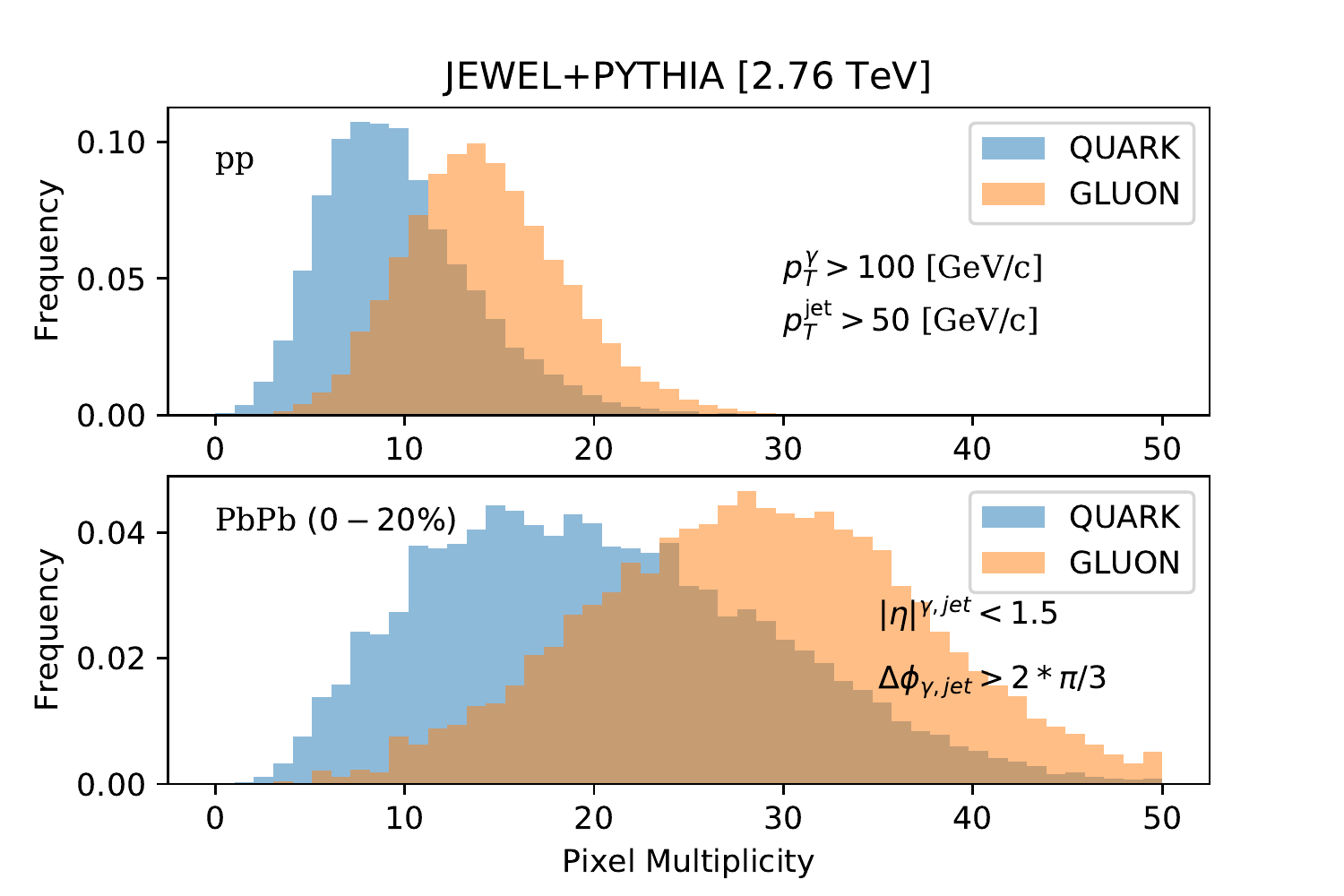}
    \includegraphics[height=3.92cm, trim = {0mm 0mm 0mm 0cm}, clip]{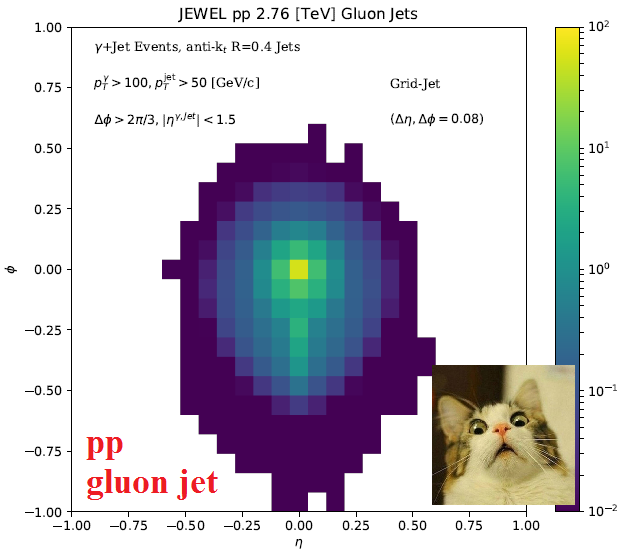}
    \includegraphics[height=3.92cm, trim = {0mm 0mm 0mm 0cm}, clip]{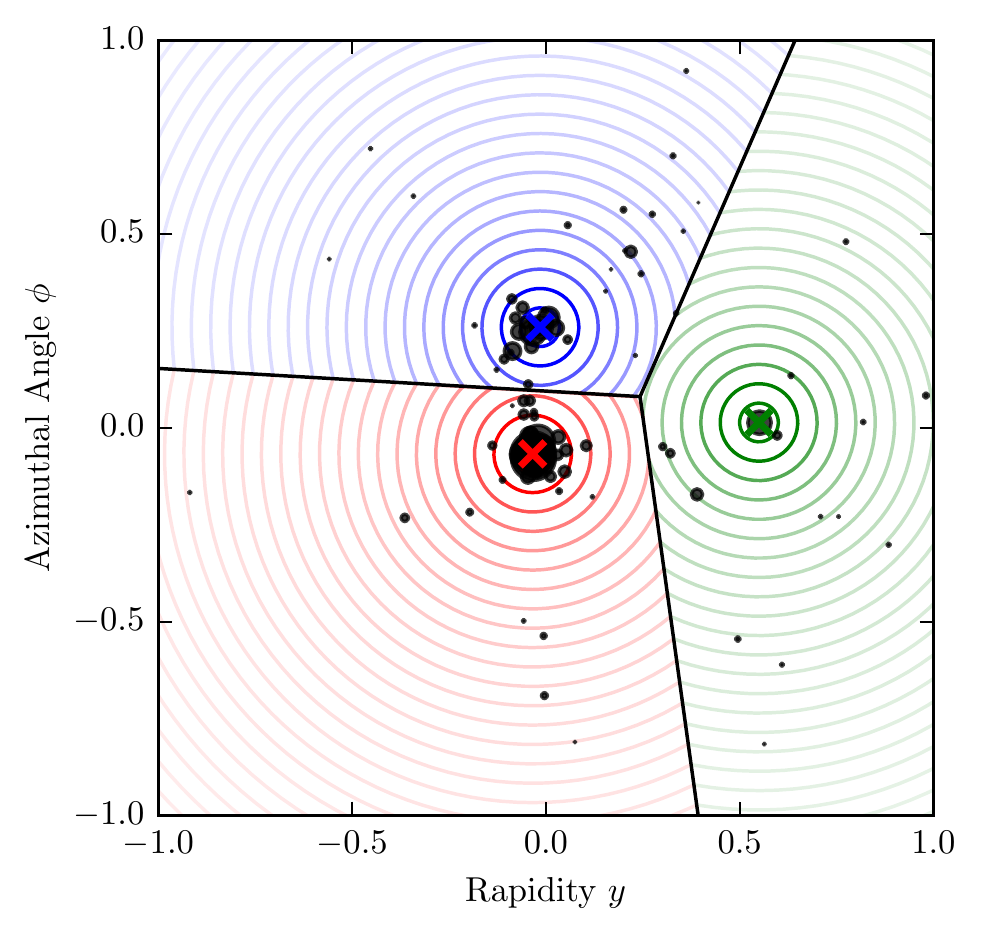}
\caption{Left panel: Pixel multiplicity distributions for quark and gluon jets in $pp$ (upper) and $AA$ (lower) collisions simulated using \textsc{Jewel}. Middle panel: Average gluon jet image in $pp$ collisions. Right panel: Telescoping deconstruction of a QCD jet at the T3 order.}
\label{fig:three}
\end{figure}

In this proceeding we exploit this idea and study classifications of quark and gluon jets in $pp$ and $AA$ collisions. The goal is to extract complete jet features which encode all aspects of jet modifications in $AA$ collisions. We study the discrimination of jets in $pp$ and $AA$ collisions and show that it is intimately related to quark-gluon discrimination (right panel of Fig. \ref{fig:sub_scale}) which aims to identify differences between quark and gluon jets. We use three approaches, starting from a multivariate analysis of a list of physics-motivated jet observables (left panel of Fig. \ref{fig:three}). On the other hand, we apply image recognition techniques which identify relevant features using machine learning methods (middle panel of Fig. \ref{fig:three}). In between, we introduce the telescoping deconstruction framework which aims to organize and capture complete physical information within jets using telescoping subjets (right panel of Fig. \ref{fig:three}). Below we briefly summarize each of the method.

\section{Quark and gluon jet substructure and modification}
\label{qgsub}

The quark and gluon enriched jet samples used in this work were generated using the prompt photon production channels $q+\gamma$ and $g+\gamma$ in \textsc{Jewel}. The physics-motivated, multivariate analysis combines information captured in each individual jet observable. We consider five representative ones: jet mass, radial moments, $p_T^D$ and pixel multiplicity. We see that gluon jets have broader energy distributions and softer hadron fragmentation compared to quark jets, and medium interactions result in broader energy distribution and softer hadron fragmentation for both quark and gluon jets. The jet image method trains a deep convolutional neural network (CNN) on quark and gluon jet images in $pp$ and $AA$ collisions \cite{Komiske:2016rsd}. The energy distribution in rapidity $y$ and azimuthal angle $\phi$ is discretized with a finite pixel size. The CNN is then capable of processing raw pixel jet data and finding useful features which help maximize the separation among jet samples. From the average jet images, we see again that gluon jets are more spread out and populating more pixels with soft particles compared to quark jets, and the medium further broadens the energy distribution.

\begin{figure}
    \centering
    \includegraphics[width=.42\columnwidth, trim = {0mm 0mm 0mm 0cm}, clip]{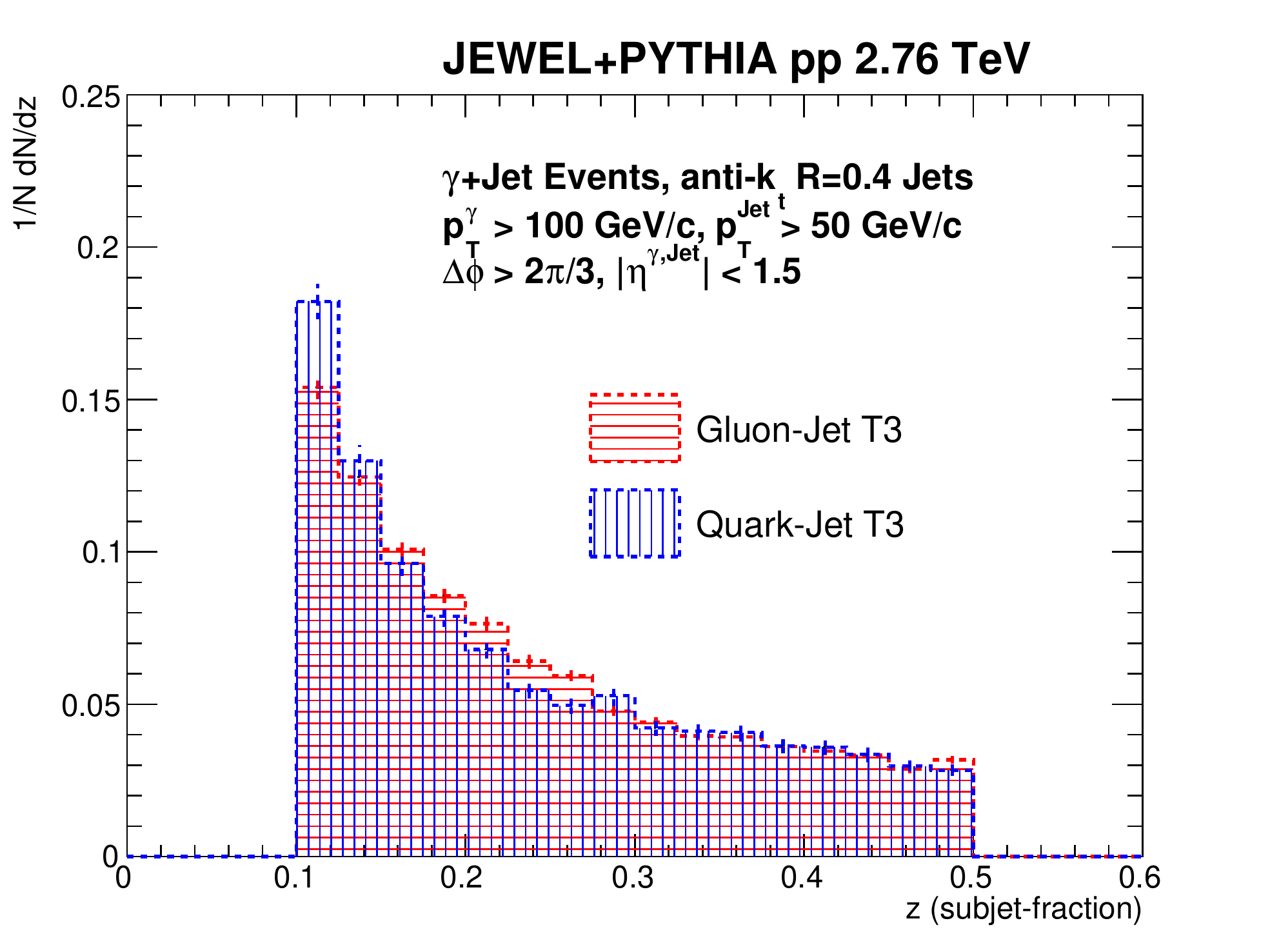}
    \includegraphics[width=.42\columnwidth, trim = {0mm 0mm 0mm 0cm}, clip]{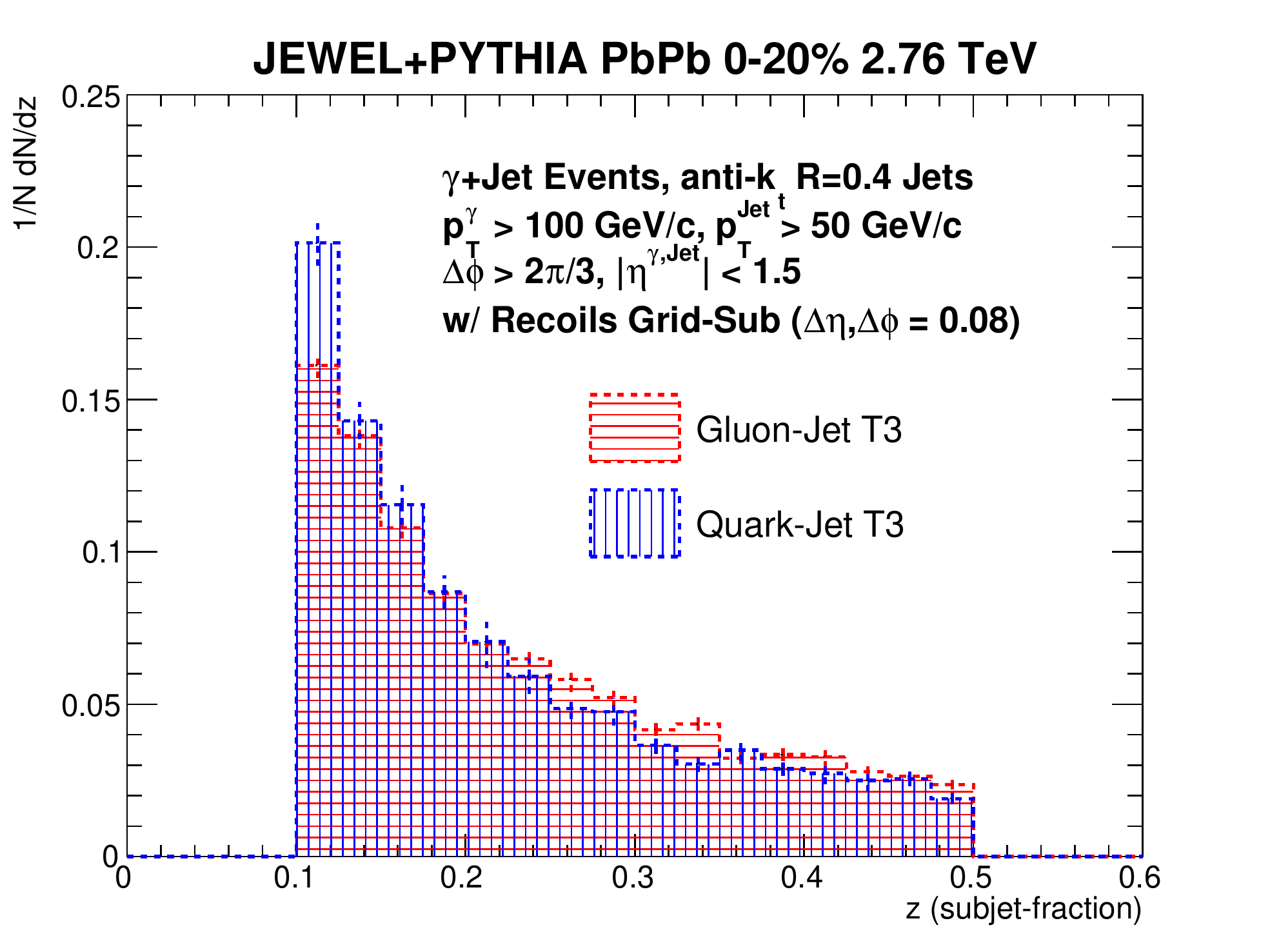}\\
    \includegraphics[width=.42\columnwidth, trim = 0mm 0cm 0mm 0cm , clip=true]{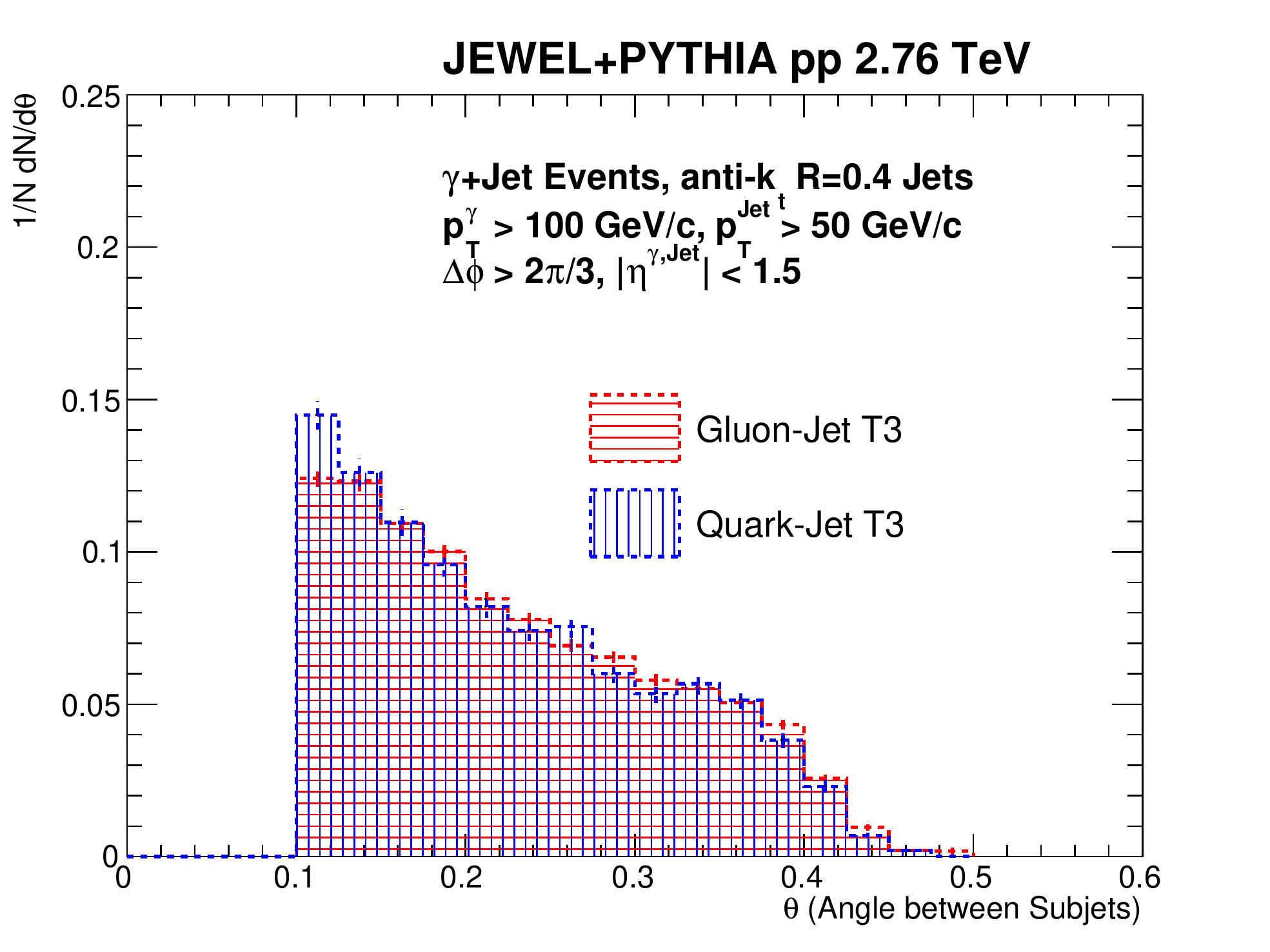}
    \includegraphics[width=.42\columnwidth, trim = 0mm 0cm 0mm 0cm , clip=true]{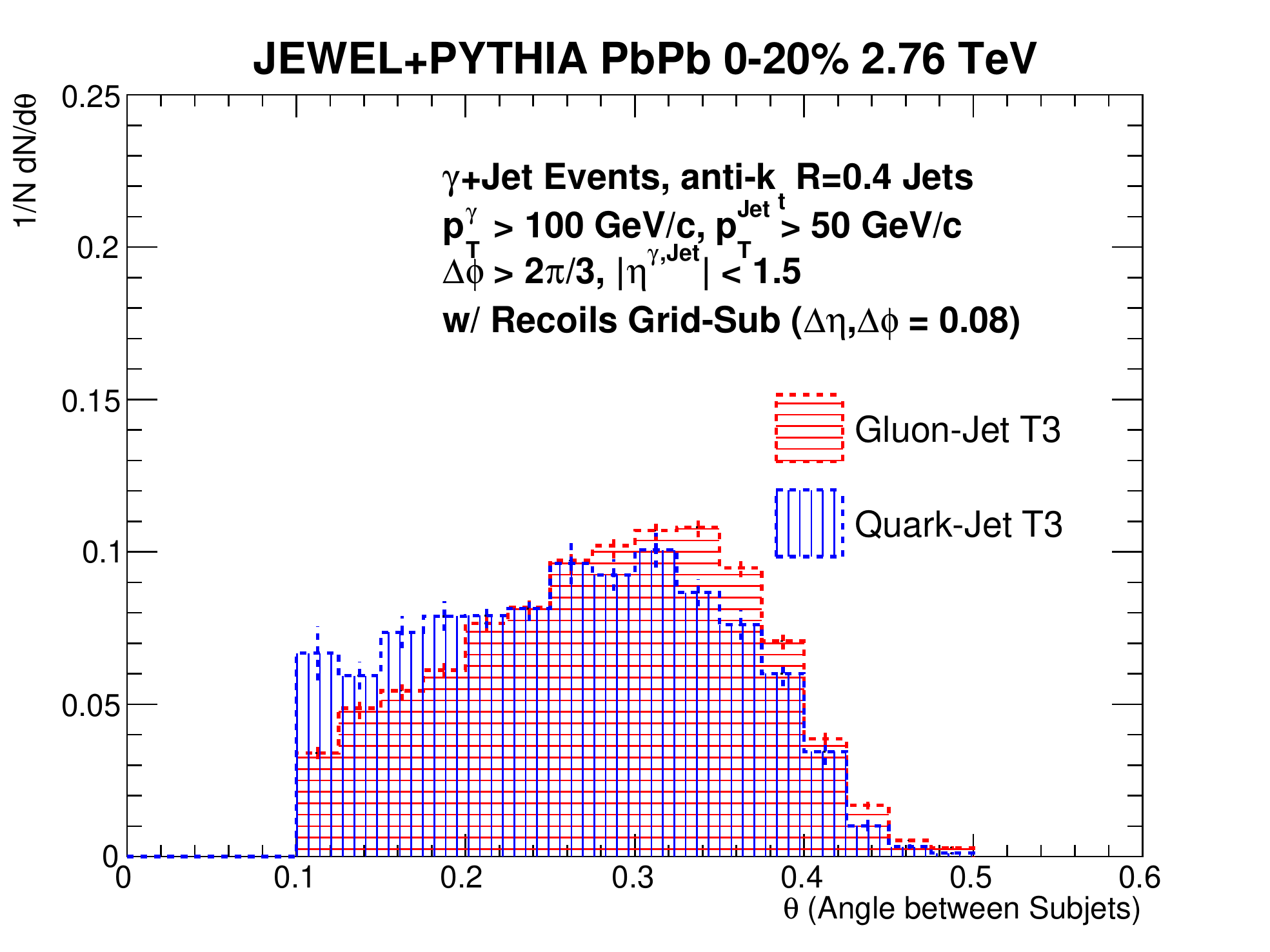}\\
    \includegraphics[width=.42\columnwidth, trim = {0mm 0mm 0mm 0cm}, clip]{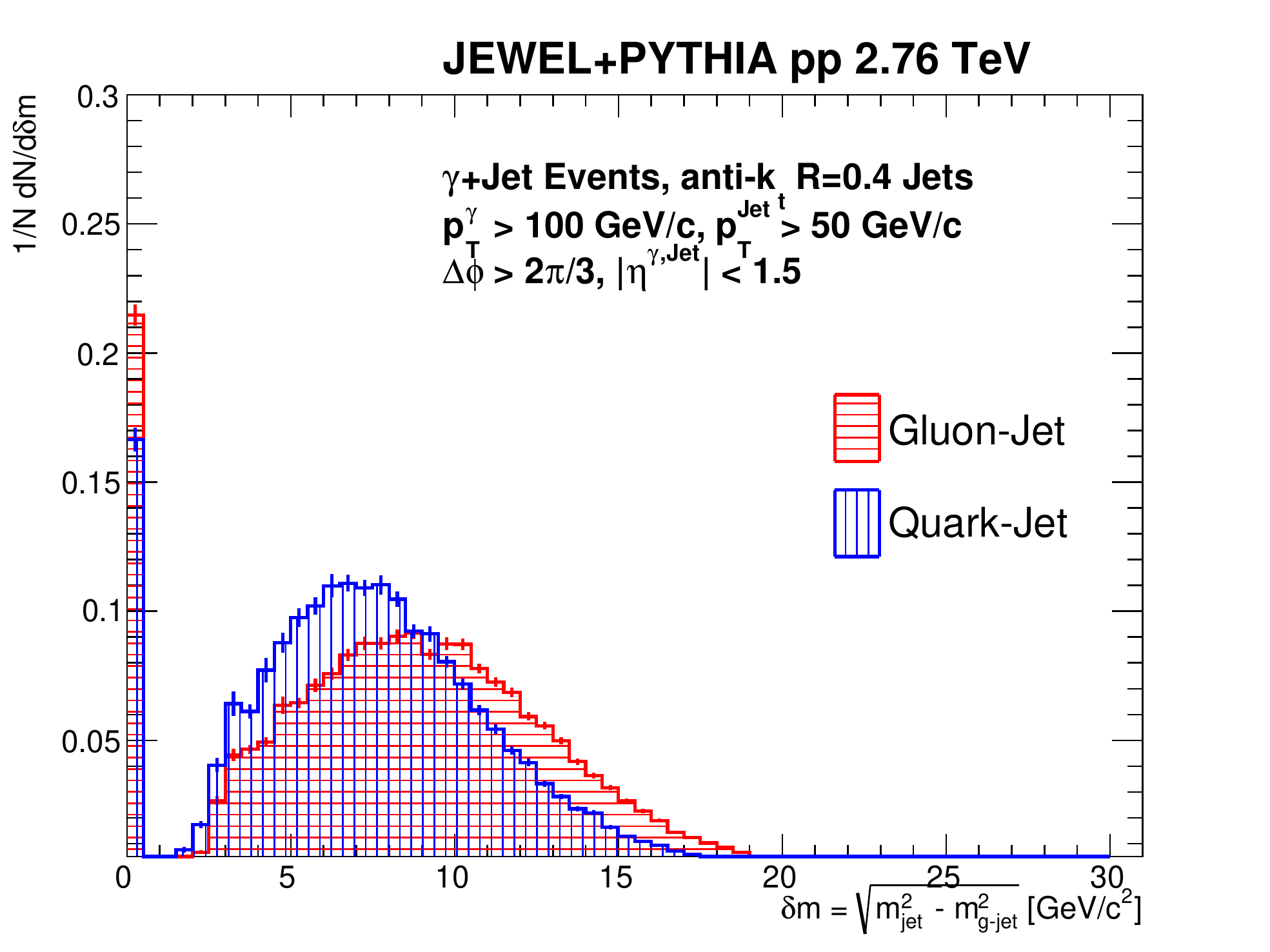}
    \includegraphics[width=.42\columnwidth, trim = {0mm 0mm 0mm 0cm}, clip]{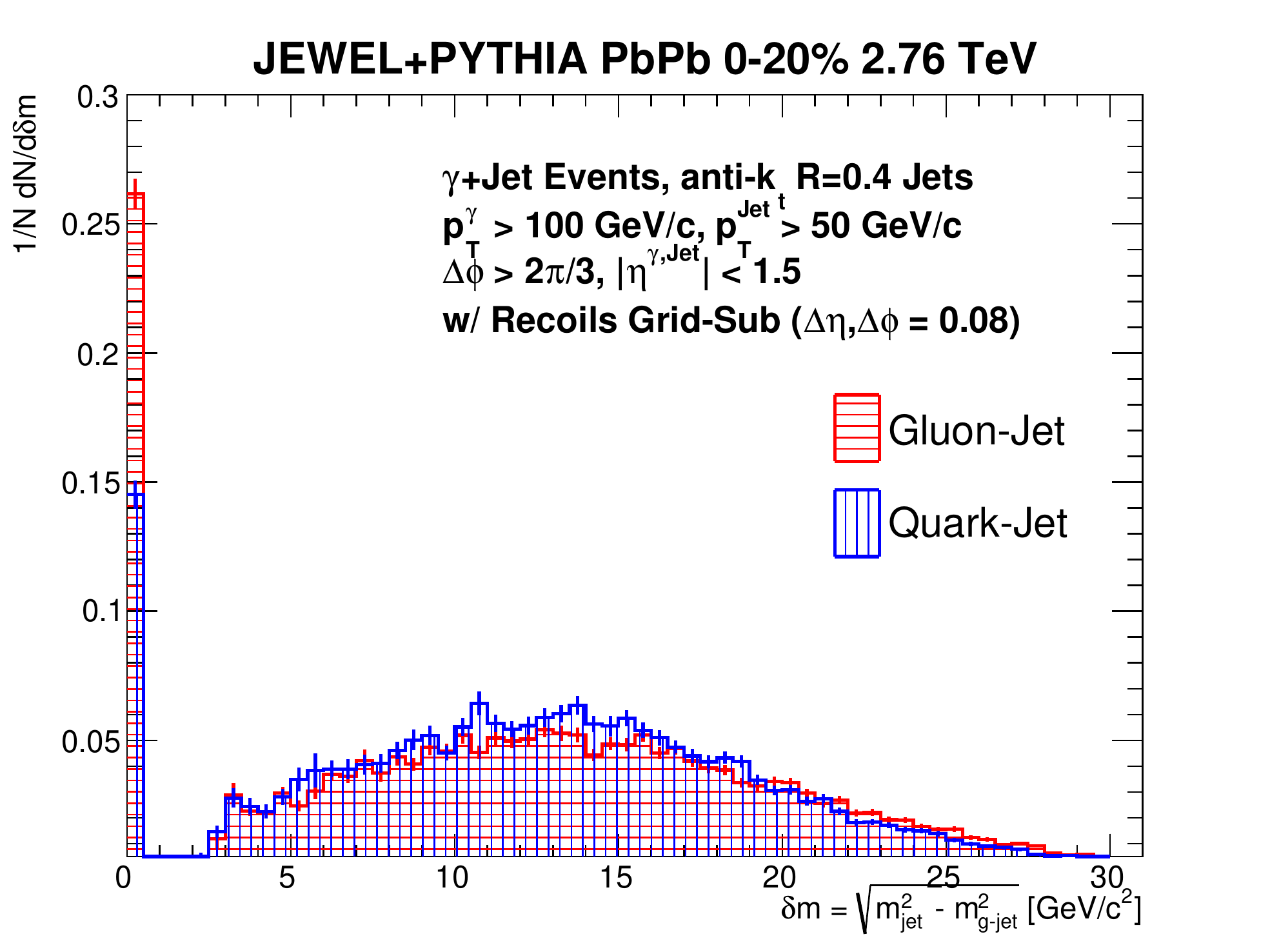}
\caption{Upper panels: Subjet momentum fraction distributions at the T3 order for quark and gluon jets in $pp$ (left) and $AA$ (right) collisions. Middle panels: Subjet angular distributions at the T3 order for quark and gluon jets in $pp$ (left) and $AA$ (right) collisions. Lower panels: Distributions of the collinear-drop observable $\delta m$ for quark and gluon jets in $pp$ (left) and $AA$ (right) collisions}
\label{fig:z_theta_cd}
\end{figure}

The TD framework probes energy flows within jets using subjets with multiple angular resolutions \cite{Chien:2013kca,Chien:2017xrb}. It decomposes jet information in a fixed-order expansion organized by the number of reconstructed subjets. At the T$N$ order, the procedure starts from identifying $N$ dominant energy flow directions along soft-recoil-free axes. Subjets are then reconstructed around the axes with multiple subjet radii $R_T$, and subjet kinematic variables form a complete jet substructure basis. We show that subjet momentum fraction $z$ and angular distributions $\theta$ constructed in telescoping deconstruction encode fundamental QCD properties such as the Altarelli-Parisi splitting functions (Fig. \ref{fig:z_theta_cd}), similar to the groomed momentum sharing $z_g$ and groomed jet radius $r_g$ constructed in Soft Drop \cite{Larkoski:2014wba}. Note the characteristic $1/z$ functional form in the subjet momentum fraction. Recently the soft-drop $z_g$ variable was used to probe heavy ion collisions with significant enhancement of soft subjets \cite{Sirunyan:2017bsd} which was first explained as a signature of medium-induced radiation \cite{Chien:2016led}. We see a similar modification pattern in the $z$ distribution and also show that the $\theta$ distribution receives strong medium modifications enhancing wide-angle emissions. To go beyond, we examine subjet masses which reveal the flavor origin of quark and gluon jets, with significant modification in $AA$ collisions. This is further tested using a collinear-drop observable $\delta m$ which is designed to probe soft radiation within jets (lower panels of Fig. \ref{fig:z_theta_cd}). We see that the difference between quark and gluon jets in $\delta m$ disappears in $AA$ collisions, which suggests that soft radiation washes out such feature that distinguishes quark and gluon jets, a possible signature of medium response to jets.

\section{Quark and gluon jet classification}
\label{qgdis}

\begin{figure}
    \centering
    \includegraphics[width=.4\columnwidth, trim = {0mm 0mm 0mm 0cm}, clip]{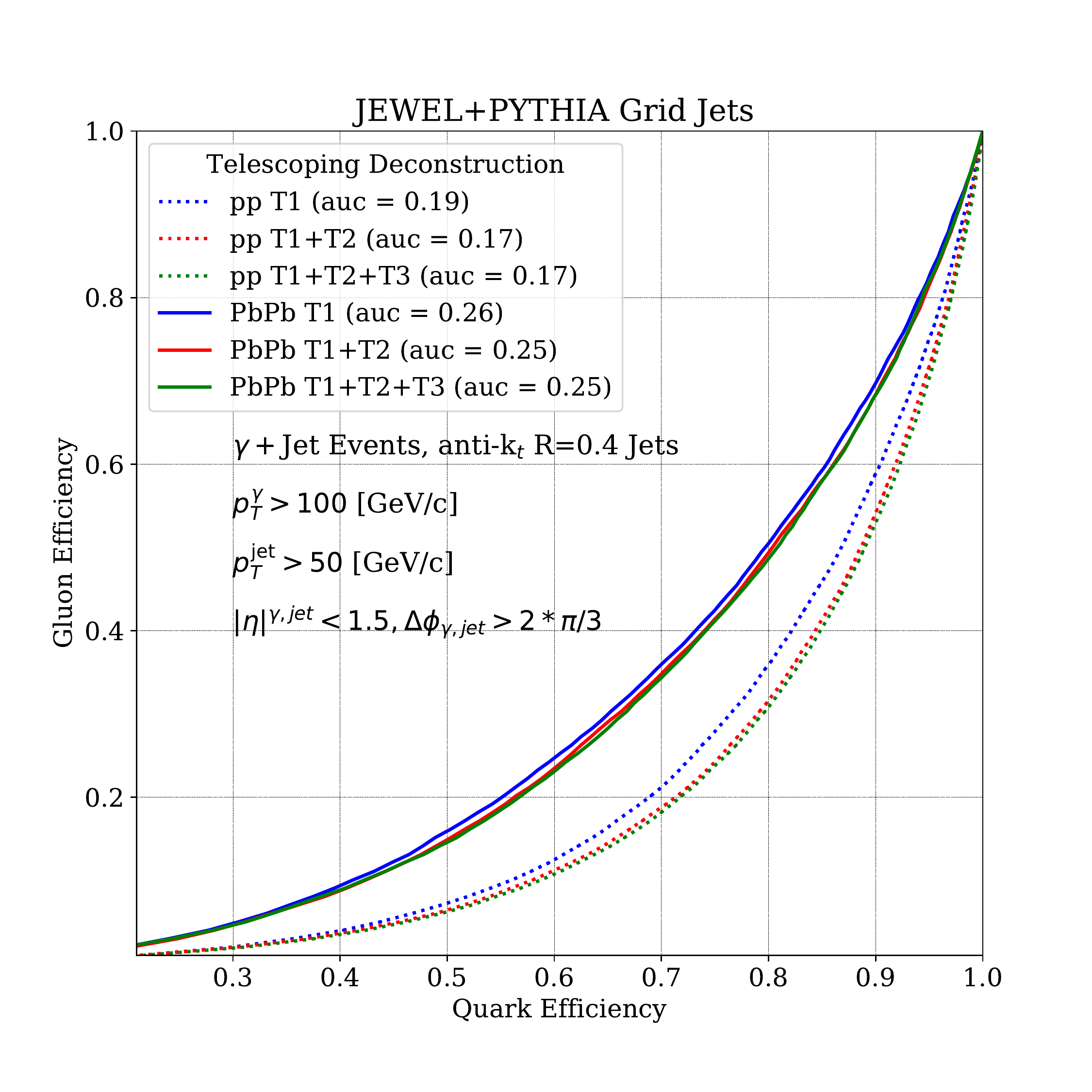}~~~~~
    \includegraphics[width=.4\columnwidth, trim = {0mm 0mm 0mm 0cm}, clip]{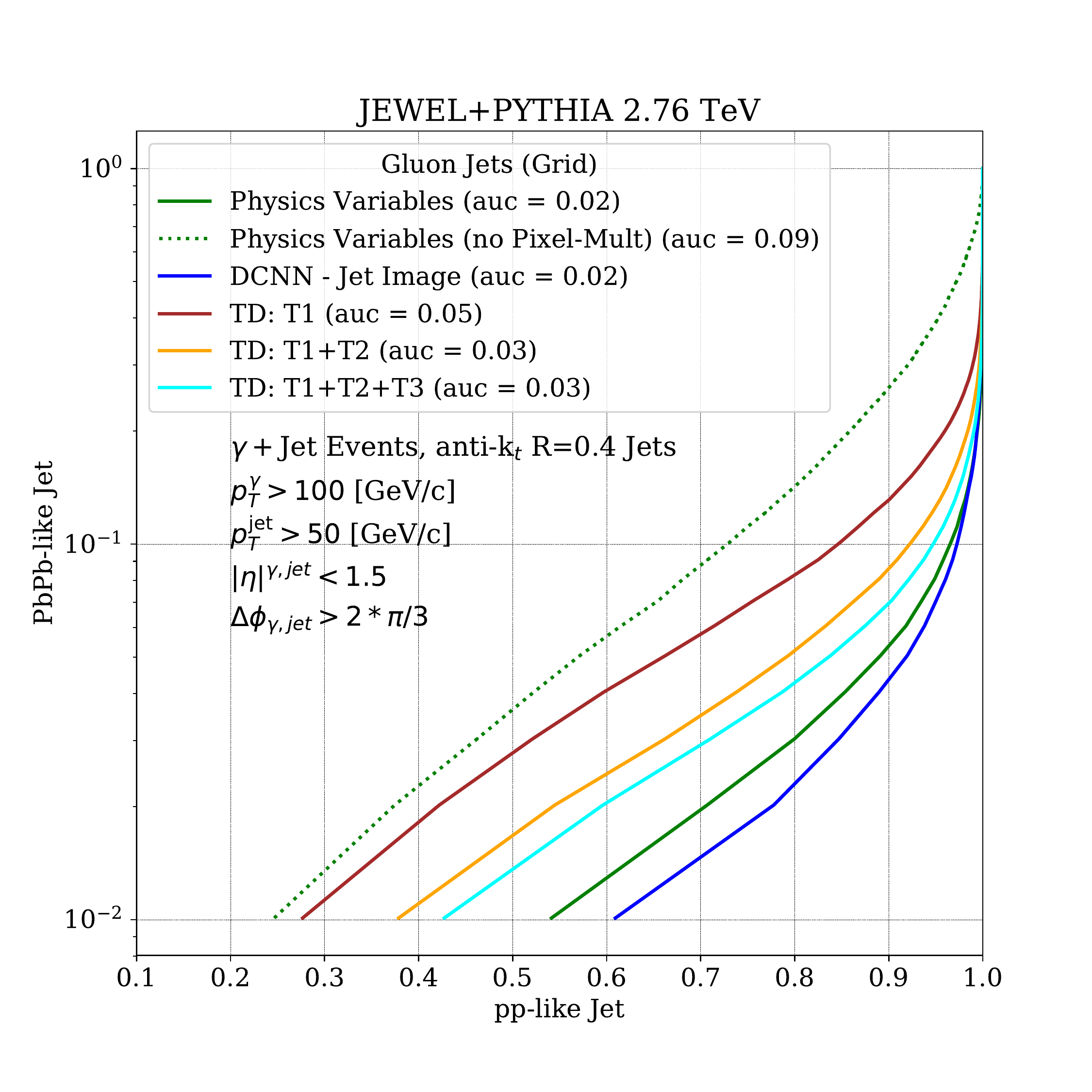}
\caption{Left panel: ROC curves using TD variables for quark-gluon discrimination in $pp$ and $AA$ collisions. Right panel: ROC curves using physics-motivated multivariate analysis, jet image and TD for discriminating gluon jets in $pp$ and $AA$ collisions. }
\label{fig:roc}
\end{figure}

Having examined the jet substructure information represented using physics-motivated observables, jet images and TD basis, we combine all the information in each category using multivariate analysis tools and study the classification of quark and gluon jets in $pp$ and $AA$ collisions. A proper neural network architecture is chosen for processing the simulated input data. We perform two tasks, discriminating quark jets and gluon jets, and discriminating jets in $pp$ and $AA$ collisions. We quantify the classification performance using receiver-operating-characteristic (ROC) curves, plotting signal efficiency versus background efficiency (Fig. \ref{fig:roc}) with higher performance towards the lower-right corner of the plots. We see that all methods give consistent and comparable performance, suggesting that with the discretization resolution each method captures most of the substructure information. The telescoping deconstruction performance converges quickly with increasing T$N$ order. We find that the quark-gluon discrimination performance goes down in \textsc{Jewel}-simulated $AA$ collisions. Also, the pixel multiplicity is the dominant observable distinguishing jets in $pp$ and $AA$ collisions, a characteristic feature of the significant soft event activities.

\section{Conclusions}
\label{conc}

We show that quark and gluon jet substructure can be independent probes of jet-medium interaction and that quark-gluon discrimination is a new way for jet modification studies. We use physics-motivated multivariate analysis and machine learning tools, and we develop the TD framework to decompose jet information using subjet basis. We emphasize that comprehensive substructure studies can lead to the understanding of the inner working of QGP.

\end{document}


\begin{figure}
\centering
    \includegraphics[height=4.2cm, trim = {0mm 0mm 0mm 0cm}, clip]{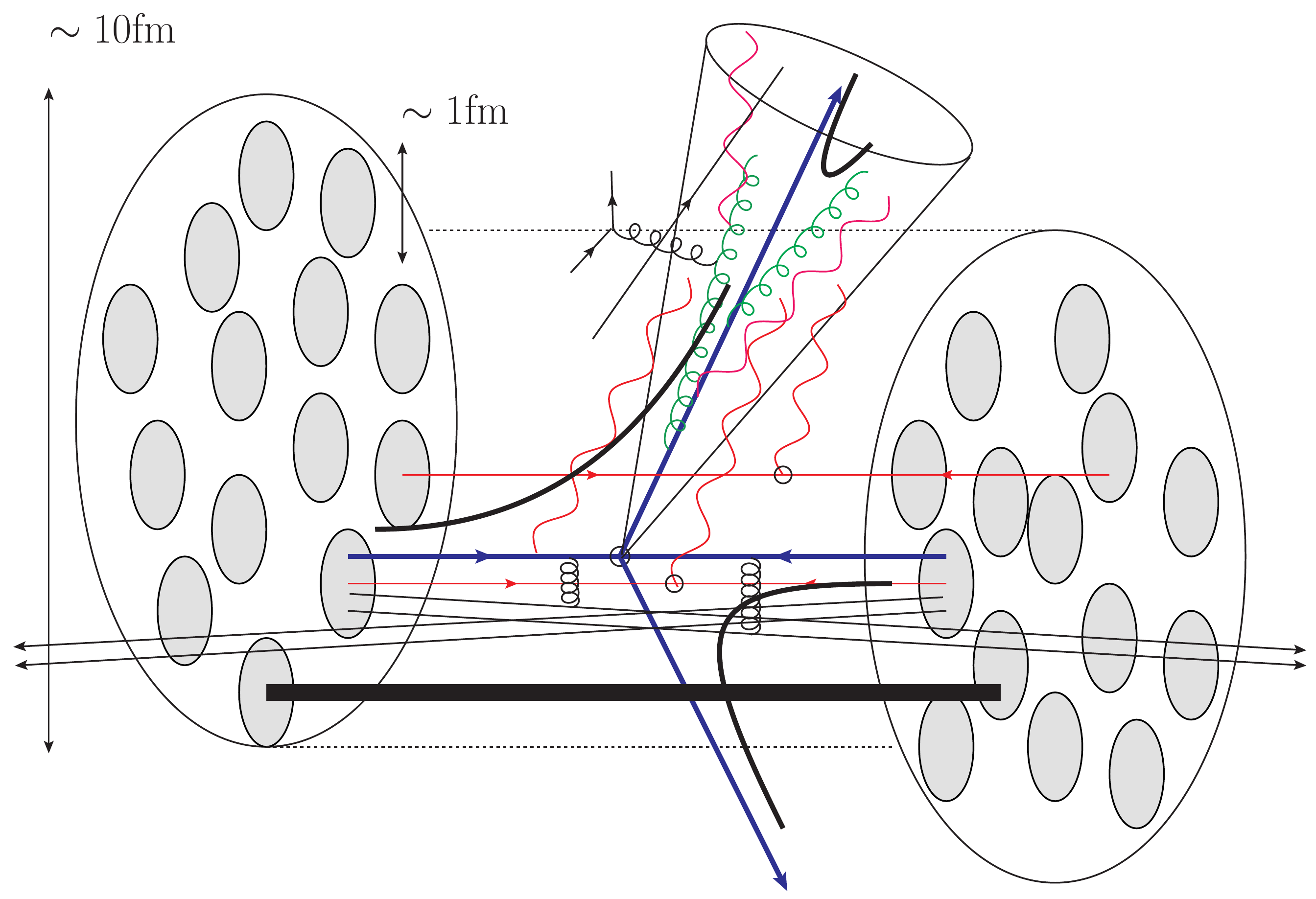}~~~~~~
    \includegraphics[height=4.2cm, trim = {0mm 0mm 0mm 0cm}, clip]{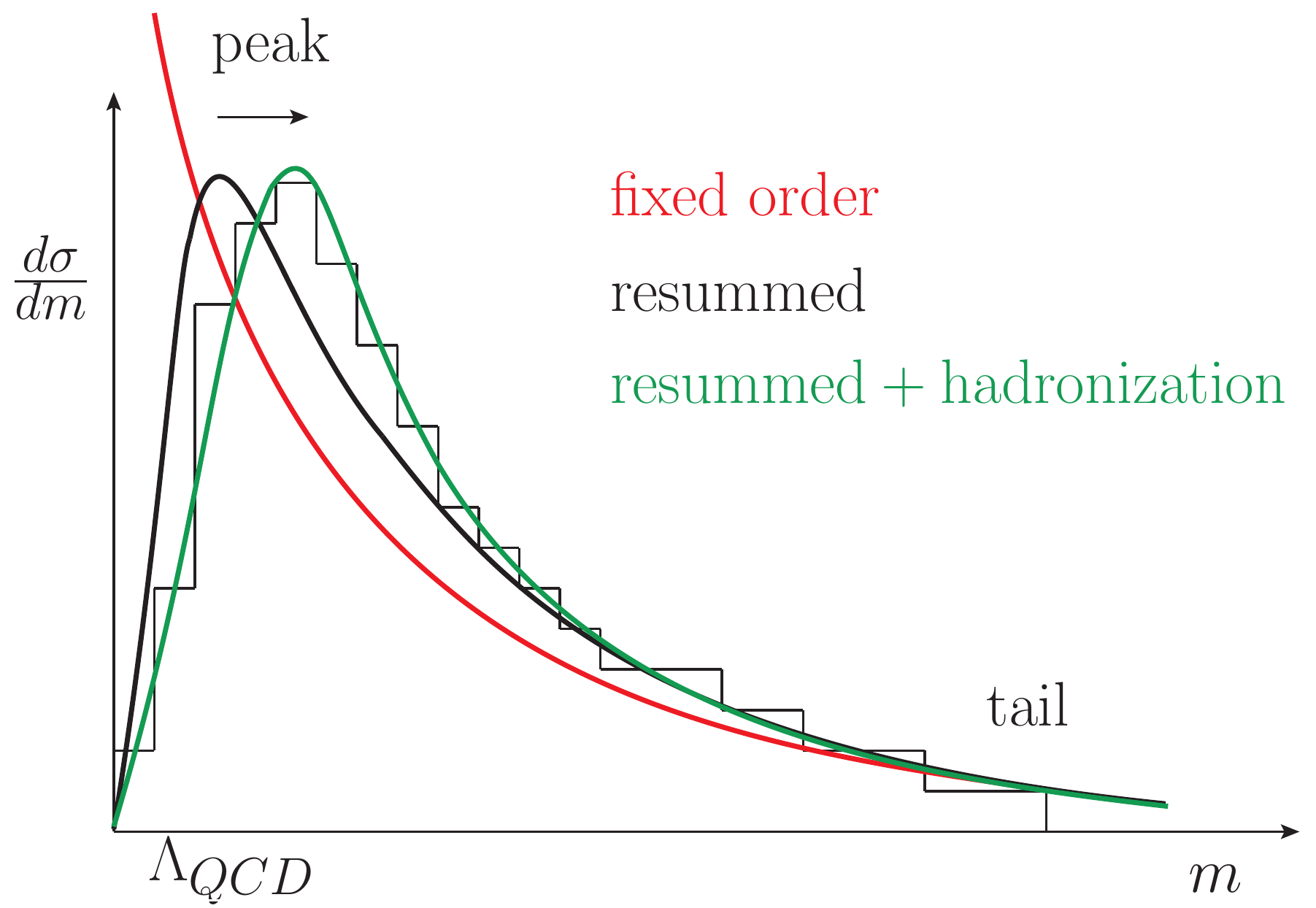}
\caption{Left panel: Various radiation contributions inside a jet in a heavy ion collision. Right panel: The necessity of all-order resummation and the effect of hadronization to the jet mass distribution. }
\label{fig:obs}
\end{figure}

\section{Introduction}
\label{intro}

A precise understanding of the redistribution of jet energy is among the priorities of jet physics in heavy ion collisions. Various proposed jet quenching mechanisms capture the picture of energy loss by transporting the jet energy to wider angle. Possible contributions from medium response to jets also affect the energy distribution. Therefore designing jet substructure observables which can probe aspects of jet-medium interaction is vital for revealing possible, novel signatures of jet quenching. On the other hand, techniques which help clean up jets -- generally referred to as jet grooming -- have been introduced to mitigate background soft radiation in order to enhance the signal sensitivity. Groomed jet observables with different grooming parameters thus have different sensitivities to radiation inside jets, giving the opportunities to test the details of jet quenching mechanisms and rigorously determine the medium properties.

\begin{figure}
\centering
    \includegraphics[height=4.3cm, trim = {0mm 0mm 0mm 0cm}, clip]{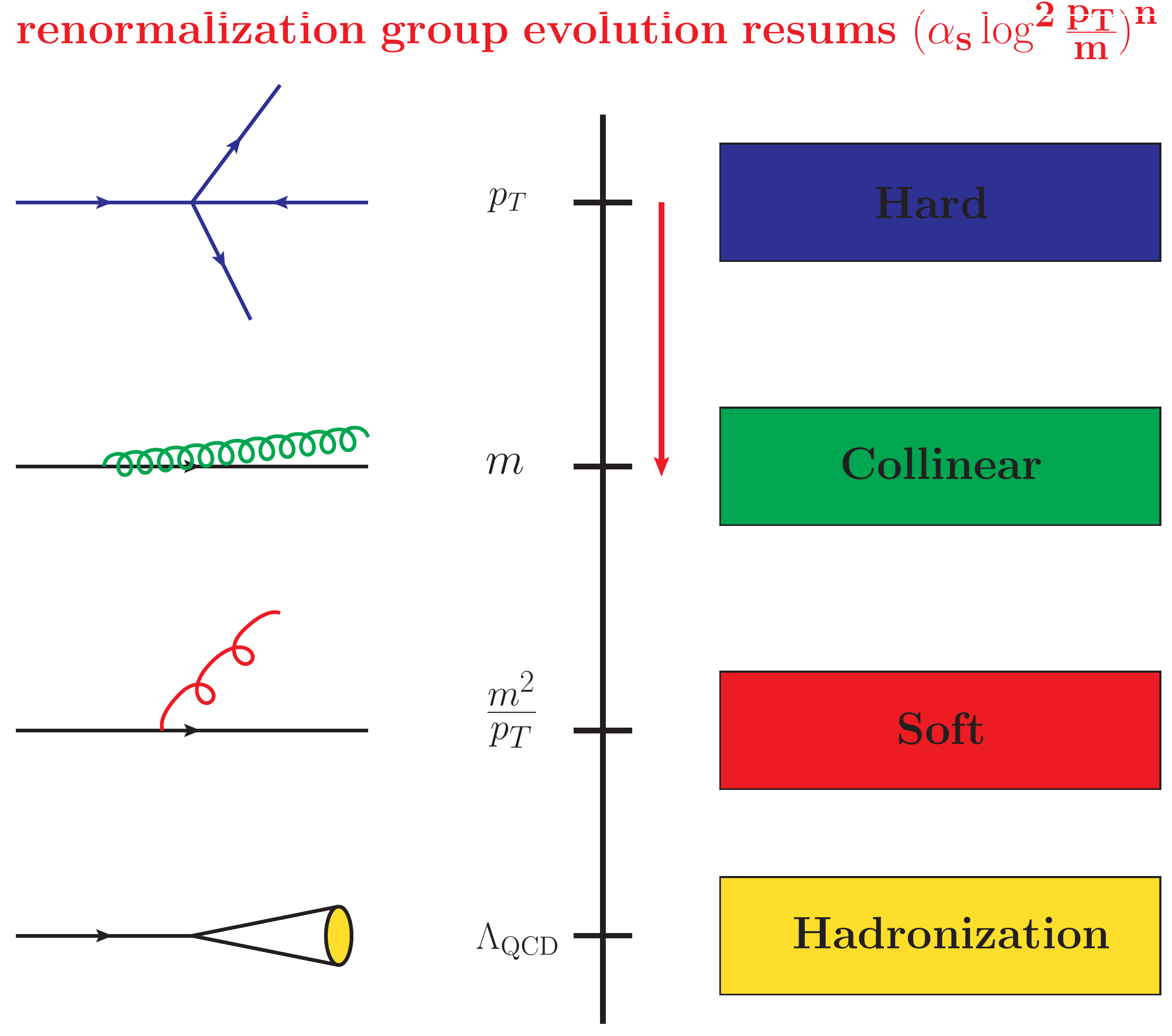}~~~~~~
    \includegraphics[height=3.3cm, trim = {0mm 0mm 0mm 0cm}, clip]{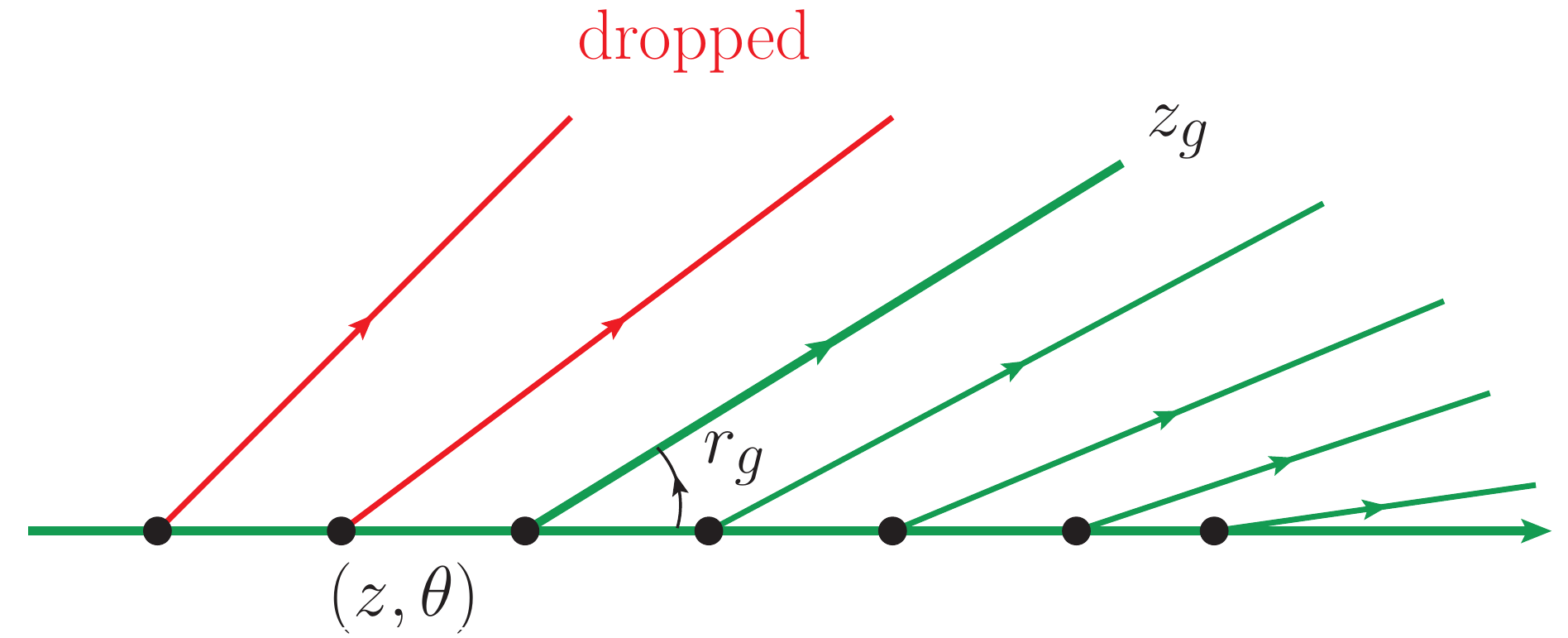}
\caption{Left panel: Factorization of jet mass in SCET and resummation performed using renormalization group evolution. Right panel: The angular-ordered tree in the soft drop procedure. }
\label{fig:framework}
\end{figure}

In this proceeding we exploit the connection between jet quenching and jet grooming and present a calculation for groomed jet mass distributions. The jet mass is a classic jet observable which is sensitive to both energetic, collinear radiation as well as soft, wide-angle radiation (represented by the green and red lines, respectively, in the left panel of Fig. \ref{fig:obs}). There are many possible sources of soft radiation, including final-state and initial-state radiation, underlying events activity correlated or uncorrelated with the hard collision, medium-induced radiation and medium response. The calculation of jet mass distribution requires all-order resummation of logarithmically-enhanced terms, as well as modeling non-perturbative contributions (right panel of Fig. \ref{fig:obs}). We use soft-collinear effective theory (SCET) to factorize the cross section, and to resum, via the renormalization group evolution, the logarithmic corrections appearing among different scales (left panel of Fig. \ref{fig:framework}). We focus on the medium modification contribution caused by induced radiation via Glauber interaction \cite{Ovanesyan:2011xy}. We use the soft drop \cite{Larkoski:2014wba} grooming procedure to remove soft, wide-angle radiation (right panel of Fig. \ref{fig:framework}). Below we summarize the formalism for calculating the soft drop jet mass distribution and present some preliminary results.

\section{Framework}
\label{formula}

Following the framework developed in \cite{Li:2012bw,Chien:2012ur,Chien:2014nsa,Chien:2015hda} for various jet substructure calculations, at next-to-leading logarithmic (NLL) accuracy one can define a process-independent groomed jet mass function $J^{SD}(m^2,\mu)$ which captures all the soft-collinear radiation inside jets ($i=q,g$ labeling quark-initiated or gluon-initiated jets),
\begin{equation}
    J_i^{SD}(m^2,\mu) = \int dp^2 dk J_i(p^2,\mu)S^{SD}_i(k,R,z_{cut},\mu)\delta (m^2-p^2-2E_J k)\;,
\end{equation}
where $S^{SD}_i(k,R,z_{cut},\mu)=S^C_i(k,R,z_{cut},\mu) S^{G}_i(R,z_{cut},\mu)$ can be further factorized. At fixed-order, the jet mass function can be directly calculated using the collinear splitting function,
\begin{equation}
    J^{SD}_i(m^2,\mu)=\sum_{j,k}\int_{PS}dxdk_\perp {\cal P}_{i\rightarrow jk}(x,k_\perp)\delta(m^2-M^2(x,k_\perp))\Theta_{\rm alg.}\Theta_{SD}\;.
\end{equation}
Here $M^2(x,k_\perp)=\frac{k^2_\perp}{x(1-x)}$ is the jet mass measurement function, while $\Theta_{\rm k_T}=\Theta(E_J R x(1-x)-k_\perp)$ and $\Theta_{SD}=\Theta(E_J R x(1-x)\Big(\frac{x}{z_{cut}}\Big)^{1/\beta}-k_\perp)$ are the jet algorithm and soft drop phase space constraints, respectively. The medium modification can also be included using medium-induced splitting functions. The full jet mass distribution can then be calculated by weighing the groomed jet mass functions with the corresponding jet cross sections,
\begin{equation}
     \frac{d\sigma}{dm^2}=\sum_{i=q,g}\int_{PS}dp_Tdy\frac{d\sigma^i}{dp_Tdy}P^{SD}_i(m^2,\mu),~{\rm where}~
     P^{SD}_i(m^2,\mu) = \frac{J^{SD}_i(m^2,\mu)}{J^i_{un}(\mu)}\;.
\end{equation}
Note that the ratio $P^{SD}_i(m^2,\mu)$ between the jet mass function and the unmeasured jet function is renormalization-group invariant and can be interpreted as the probability distribution of the jet mass for jets initiated by parton $i$.

\section{Preliminary results}
\label{result}

\begin{figure}
\centering
    \includegraphics[height=4cm, trim = {0mm 0mm 0mm 0cm}, clip]{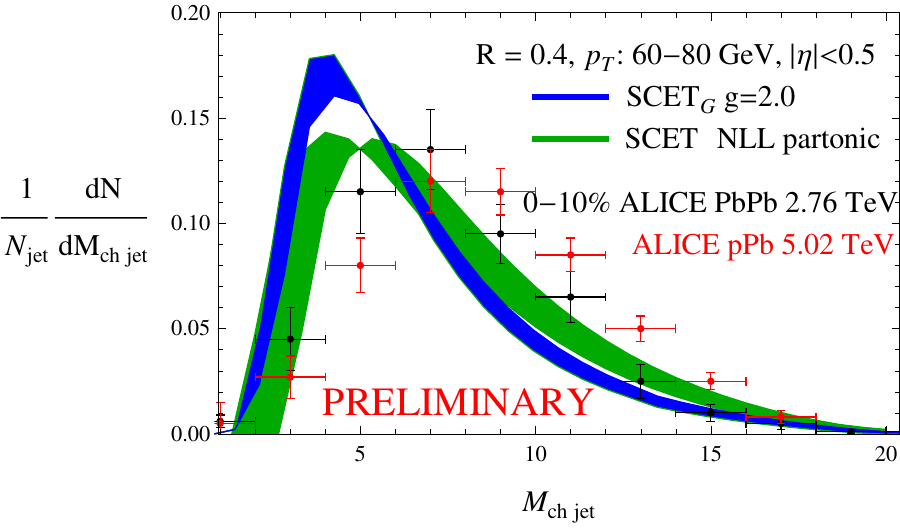}~~~~~~
    \includegraphics[height=4cm, trim = {0mm 0mm 0mm 0cm}, clip]{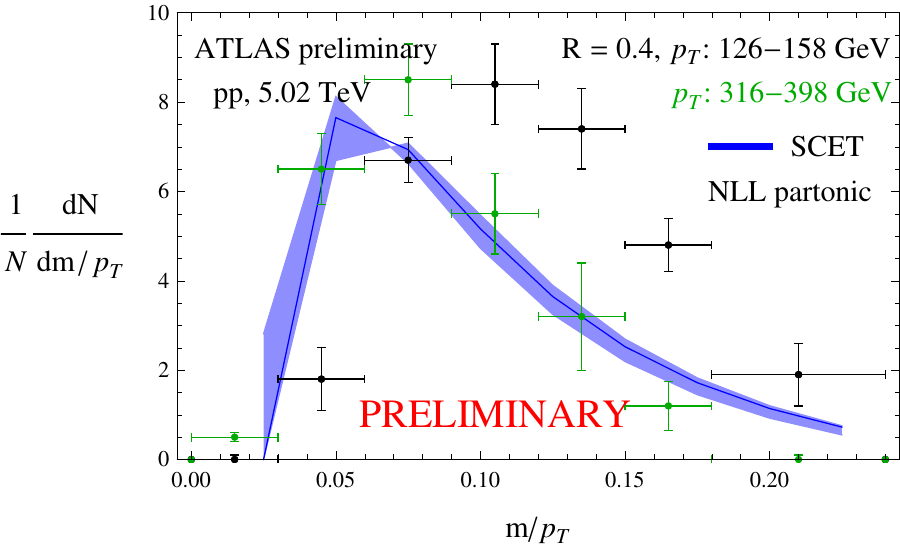}
\caption{Left panel: NLL SCET calculations (assuming isospin symmetry) and ALICE measurements \cite{Acharya:2017goa} of inclusive charged jet mass distributions in 2.76 TeV PbPb and 5.02 TeV pPb collisions. Right panel: NLL SCET calculation and ATLAS measurement \cite{ATLAS:2018jsv} of inclusive jet mass over transverse momentum ratio in 5.02 TeV pp collisions.}
\label{fig:result1}
\end{figure}

Having outlined the framework of our calculation, we present some preliminary results to compare with PYTHIA simulations and to currently available measurements.  The left panel of Fig. \ref{fig:result1} shows the comparison between NLL SCET calculations and ALICE measurements \cite{Acharya:2017goa} of inclusive charged jet mass distributions in 2.76 TeV PbPb and 5.02 TeV pPb collisions. In the calculation we relate the total jet mass to the charged jet mass assuming isospin symmetry. The calculation does not yet include non-perturbative effects therefore the jet mass is systematically lower than the measured values. However, we can see that the medium modifies the jet mass towards smaller values. The right panel of Fig. \ref{fig:result1} shows the comparison between SCET calculation and ATLAS measurement \cite{ATLAS:2018jsv} of inclusive jet mass over jet transverse momentum ($m/p_T$) distribution in 5.02 TeV pp collisions. We see that the partonic calculation agrees well with the measured distribution for jets at higher $p_T$ where non-perturbative effects are small.

\begin{figure}
\centering
    \includegraphics[height=4cm, trim = {0mm 0mm 0mm 0cm}, clip]{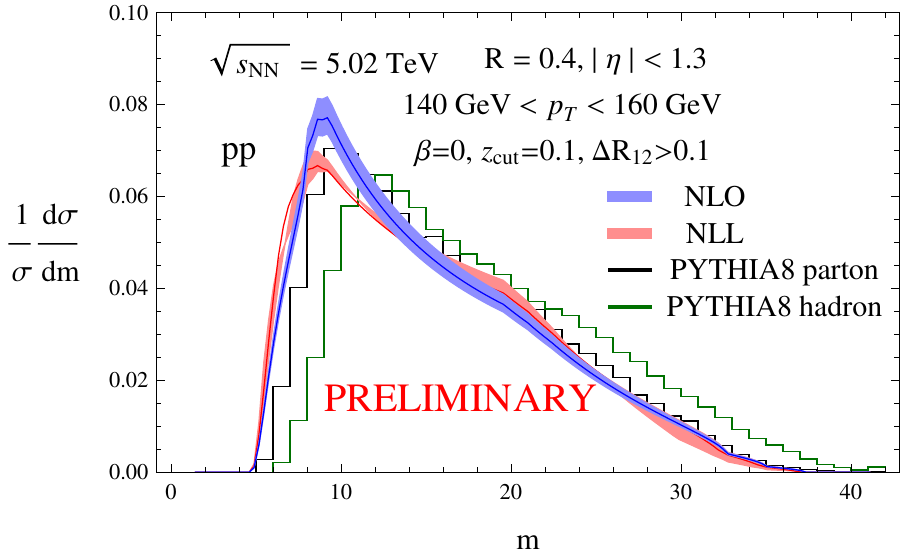}~~~~~~
    \includegraphics[height=4cm, trim = {0mm 0mm 0mm 0cm}, clip]{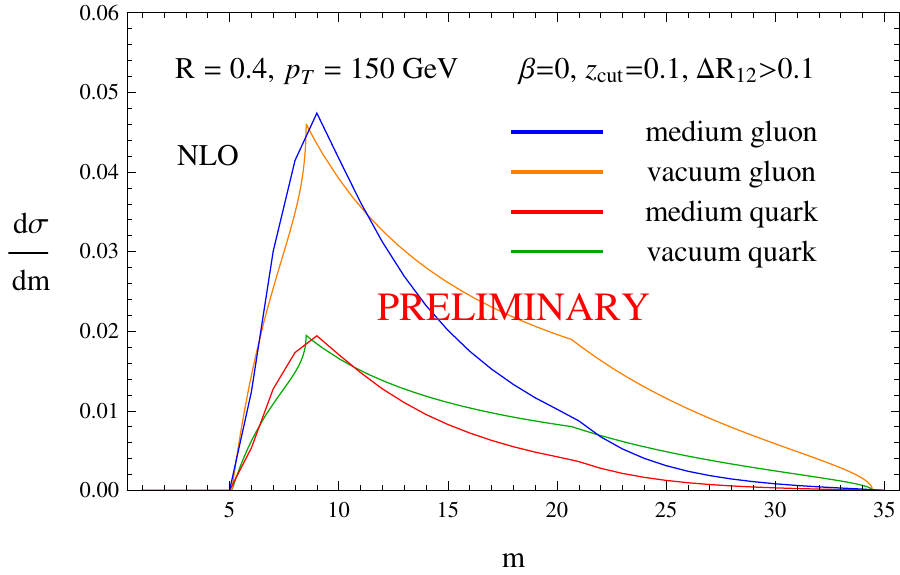}
\caption{Left panel: PYTHIA simulations compared to the NLO and NLL+NLO SCET calculations of inclusive soft drop jet mass distribution in 5.02 TeV pp collisions. Right panel: NLO vacuum and medium contributions to inclusive soft drop jet mass distribution in SCET. }
\label{fig:result2}
\end{figure}

The left panel of Fig. \ref{fig:result2} shows the comparison of NLO and NLL+NLO SCET calculation with PYTHIA simulations at parton and hadron level for inclusive soft drop jet mass distribution in 5.02 TeV pp collisions. We see that the parton level results agree quite well, and that hadronization affects the distribution significantly in all regions. The right panel shows the NLO vacuum and medium contributions and suggests that the small jet mass region is enhanced by the medium.
 
\begin{figure}
\centering
    \includegraphics[height=4.8cm]{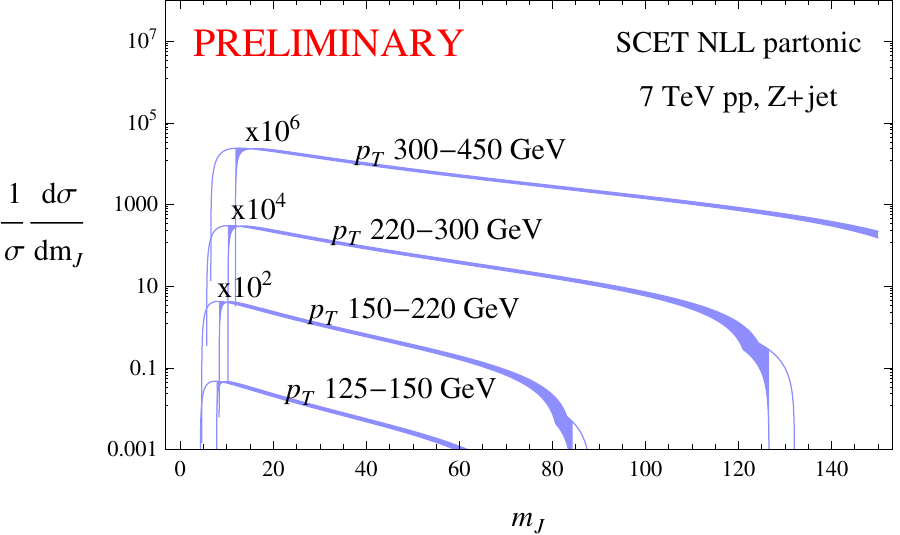}~~~~~~
    \includegraphics[height=5cm]{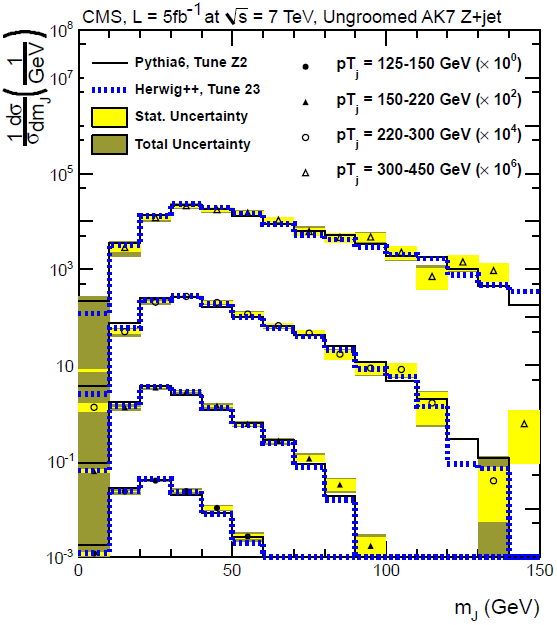}
    \caption{NLL SCET calculation (left panel) and CMS measurements \cite{Chatrchyan:2013vbb} (right panel) of jet mass distribution in $Z$+jet events in 7 TeV pp collisions. }
\label{fig:result3}
\end{figure}

The CMS measurement imposes a $\Delta R_{12}$ cut to constrain the angle between two soft drop branches therefore removing the Sudakov peak region \cite{Sirunyan:2018gct}. The formalism allows us to calculate the jet mass distribution for jets from arbitrary hard scattering process. Fig. \ref{fig:result3} shows the NLL SCET calculation (left panel) of jet mass distribution for $Z$+jet events in 7 TeV pp collisions and compare with CMS measurements \cite{Chatrchyan:2013vbb} (right panel). We see the very good agreement which further confirms the validity of the framework.

\section{Conclusions}
\label{conc}

Jet mass contains rich information of the radiation inside jets, and jet grooming gives a concrete strategy to probe specific regions which can reveal possible jet quenching features. We provide an effective field theory framework to systematically resum logarithmically-enhanced contributions and calculate the jet mass distribution at NLL accuracy for any arbitrary hard scattering process with jets in the final state. This work provides a baseline for future precision jet substructure studies.